\documentclass{Interspeech2024}
\usepackage{amsmath,amsfonts,amsthm,bm}
\usepackage{booktabs}
\usepackage{multirow}
\usepackage[table,xcdraw]{xcolor}
\DeclareMathOperator*{\argmin}{arg\,min} 
\usepackage{cite}

\renewenvironment{thebibliography}[1]{
 \begin{oldthebibliography}{#1}
 \setlength{\itemsep}{0.1em}
 \setlength{\parskip}{0.1em}
}
{
 \end{oldthebibliography}
}

\makeatletter
\def\bstctlcite{\@ifnextchar[{\@bstctlcite}{\@bstctlcite[@auxout]}}
\def\@bstctlcite[#1]#2{\@bsphack
  \@for\@citeb:=#2\do{%
    \edef\@citeb{\expandafter\@firstofone\@citeb}%
    \if@filesw\immediate\write\csname #1\endcsname{\string\citation{\@citeb}}\fi}%
  \@esphack}
\makeatother

 \interspeechcameraready

\title{Diffusion-based Generative Modeling with Discriminative Guidance for Streamable Speech Enhancement} 

\name[affiliation={1,2}]{Chenda}{Li}
\name[affiliation={2}]{Samuele }{Cornell}
\name[affiliation={2}]{Shinji}{Watanabe}
\name[affiliation={1}]{Yanmin}{Qian}

\address{
  $^1$Shanghai Jiao Tong University, China \\
  $^2$Carnegie Mellon University, USA
  }
\email{lichenda1996@sjtu.edu.cn, cornellsamuele@gmail.com, 	shinjiw@ieee.org, yanminqian@sjtu.edu.cn}


\begin{document}
\bstctlcite{IEEEexample:BSTcontrol} 
\maketitle

\begin{abstract}

Diffusion-based generative models (DGMs) have recently attracted attention in speech enhancement research (SE) as previous works showed a remarkable generalization capability. However, DGMs are also computationally intensive, as they usually require many iterations in the reverse diffusion process (RDP), making them impractical for streaming SE systems. In this paper, we propose to use \textit{discriminative scores} from discriminative models in the first steps of the RDP. These discriminative scores require only one forward pass with the discriminative model for multiple RDP steps, thus greatly reducing computations. This approach also allows for performance improvements. We show that we can trade off between generative and discriminative capabilities as the number of steps with the discriminative score increases. Furthermore, we propose a novel streamable time-domain generative model with an algorithmic latency of $50$ ms, which has no significant performance degradation compared to offline models.
\end{abstract}

\section{Introduction}

Speech enhancement (SE) \cite{loizouSpeechEnhancementTheory2007} aims at improving the quality of speech degraded by additive noise or reverberation.
In recent years, supervised deep neural networks (DNN)-based methods have drastically advanced SE techniques.  
The prevailing approach is to train a DNN with a regression objective to estimate the clean target speech from the noisy signal, using time-frequency (T-F) domain spectrum masking \cite{wangSupervisedSpeechSeparation2018,huDCCRNDeepComplex2020}, T-F domain spectrum mapping \cite{xuRegressionApproachSpeech2015,wangComplexSpectralMapping2020b} or estimating the time-domain waveform directly \cite{pandeyTCNNTemporalConvolutional2019,luoConvTasNetSurpassingIdeal2019,kinoshitaImprovingNoiseRobust2020}. 
This \textit{discriminative} (also referred to as \textit{predictive} ~\cite{pml2Book}) approach, while highly effective with in-domain data, can sometimes introduce distortions~\cite{zhangClosingGapTimeDomain2021a,wangBridgingGapMonaural2020a} especially when there is a mismatch between the simulated training data and the real-world deployment scenario. 
On the other hand, recent works hint at the possibility that \textit{generative} methods, and in particular diffusion-based generative models (DGM)~\cite{luConditionalDiffusionProbabilistic2022, welkerSpeechEnhancementScoreBased2022,richterSpeechEnhancementDereverberation2023a, lemercierStoRMDiffusionBasedStochastic2023}, could be more robust to such mismatched conditions~\cite{luConditionalDiffusionProbabilistic2022,richterSpeechEnhancementDereverberation2023a} and generally produce less intrusive artifacts than \textit{discriminative} approaches, as suggested by their superior results with perceptual metrics~\cite{welkerSpeechEnhancementScoreBased2022,richterSpeechEnhancementDereverberation2023a, lemercierStoRMDiffusionBasedStochastic2023}.



However, DGMs require a multi-step iterative update in the diffusion reverse process for inference. 
In each reverse diffusion step, the \textit{score function} DNN will perform a forward inference step at least once~\cite{songScoreBasedGenerativeModeling2021a}. 
To achieve the desired performance, the total number of reverse diffusion steps $N$ is usually large, e.g., $N=30$ in \cite{richterSpeechEnhancementDereverberation2023a} and $N=50$ in \cite{luConditionalDiffusionProbabilistic2022}.
This leads DGM to have practically a much higher computational cost than discriminative methods, all other things being equal (e.g., the DNN architecture). 
This is the most significant problem that prevents DGMs from being used for online speech enhancement.

In this paper, we follow this latter line of research and study how DGMs can be leveraged for online, streamable SE. In fact, online SE is crucial for many applications such as telecommunications \cite{dubey2023icassp}, hearing aids \cite{graetzerClarity2021ChallengesMachine2021a}, automatic speech recognition (ASR) \cite{haeb-umbachSpeechProcessingDigital2019}, etc.
Recently~\cite{richterSpeechSignalImprovement2023a,richterCausalDiffusionModels2024} also explored streamable SE with DGMs. However, this work only addressed ideal/algorithmic latency. The above-mentioned issue would remain a problem for real-world inference, where the actual computational load is the crucial factor. 
To lower the inference-time computational load, we propose, in the reverse diffusion process, to estimate the gradients of the log data distribution probability (i.e., the \textit{score function} \cite{songScoreBasedGenerativeModeling2021a}) with the guidance of an arbitrary discriminative SE model.
In the first $N_{\phi}$ steps ($N_{\phi} < N$) of the reverse process, we replace the DNN score model with a discriminative \textit{score} derived from the discriminative model enhanced speech. 
As the discriminative model only requires one forward inference step for the whole reverse diffusion process, this approach leads to significant reductions in the computational overhead.

Besides minimizing computational overhead, experimental results demonstrate our ability to obtain a flexible balance between accuracy and generalization: $N_{\phi}$ can be used to trade-off between discriminative and generative capabilities of the two SE models.
With a larger $N_{\phi}$, the reverse process is faster but also closer to the discriminative model results and vice versa. Remarkably, we show that by choosing an appropriate $N_{\phi}$, a system with the advantages of both generative and discriminative models can be achieved.


As an additional contribution of this work, to make SE with DGMs streamable, we formulate the diffusion process in the time domain in a chunk-level manner with an algorithmic latency of $50$\,ms. While this is unsuitable for strict low-latency applications such as hearing-aid SE, it is still acceptable for applications such as ASR and telecommunications, including teleconferencing~\cite{munhall1996temporal}. 
We propose a DGM for streamable SE based on the Skipping Memory (SkiM) \cite{liSkimSkippingMemory2022} model. 
SkiM is suitably modified here to perform such chunk-level diffusion process.
The history of all the previously processed chunks maintained in SkiM minimizing the performance gap between online and offline models.



\section{Methods}

\subsection{Score-based generative speech enhancement}
Most SE DGM works~\cite{welkerSpeechEnhancementScoreBased2022,richterSpeechEnhancementDereverberation2023a} can be formulated within the score-based generative modeling (SGM) through stochastic differential equations (SDEs) framework~\cite{songScoreBasedGenerativeModeling2021a}. This latter unifies discrete-time denoising diffusion probabilistic models (DDPM)~\cite{hoDenoisingDiffusionProbabilistic2020} and \textit{score matching} with Langevin dynamics~\cite{songGenerativeModelingEstimating2019} in a single framework based on a continuous-time diffusion process.
A typical SGM has three main components: a \textit{forward diffusion process} defined by an SDE, a \textit{reverse diffusion process} defined by its corresponding reverse SDE, and a \textit{numerical solver} to generate samples by solving the reverse SDE.
 
\noindent\textbf{Forward diffusion process:} recent works \cite{welkerSpeechEnhancementScoreBased2022,richterSpeechEnhancementDereverberation2023a} applied the SGM framework to the speech enhancement (SGMSE) task and defined the diffusion process in the complex spectrum domain. 
In this paper, we define the \textit{diffusion process} instead directly in the time domain, and our forward SDE can be defined as:
\begin{align}
    \mathrm{d}\mathbf{x}_t = \mathbf{f}(\mathbf{x}_t, \mathbf{y})\mathrm{d}t + g(t)\mathrm{d}\mathbf{w}, 
\end{align}
where $\mathbf{x}_t \in \mathbb{R}^{1\times L}$ is the current state of a single channel time-domain signal of length $L$ and  $\mathbf{y} \in \mathbb{R}^{1\times L}$ is the noisy speech signal. $t$ in interval $[0, T]$ is a continuous variable presenting the current $t$-step in the diffusion process $\{\mathbf{x}_t\}_{t=0}^T$. Specially, $\mathbf{x}_{0} \sim p_{0}$ is sampled from the clean speech distribution $p_{0}$, and $\mathbf{x}_{T} \sim \mathcal{N}(\mathbf{x}_{T}; \mathbf{y}, \sigma(T)^{2}\mathbf{I}) $ is sampled from the prior Gaussian distribution centered around the noisy speech $\mathbf{y}$ with variance $\sigma(T)^2$.  The vector-valued function $\mathbf{f}(\mathbf{x}_t, \mathbf{y})$ is named as \textit{drift coefficient}:
\begin{align}
\label{eq:drift} 
    \mathbf{f}(\mathbf{x}_t, \mathbf{y}) := \gamma(\mathbf{y} - \mathbf{x}_{t}),
\end{align}
where hyperparameter $\gamma$ is a positive constant called \textit{stiffness}, it controls the current-step signal $\mathbf{x}_t$ drift towards noisy speech $\mathbf{y}$. At the same time, $\mathbf{w} \in \mathbb{R}^{1 \times L} $ is a small Gaussian increment in the standard Wiener process, and  $g(t)$ is the \textit{diffusion coefficient}, it controls the scale of Gaussian increment $\mathbf{w}$ added in the current step $t$.

\noindent\textbf{Reverse diffusion process:} the \textit{reverse diffusion process} can be solved according to the reverse SDE 
\cite{andersonReversetimeDiffusionEquation1982, songScoreBasedGenerativeModeling2021a}:
\begin{align}
\label{eq:reverse_sde}
    \mathrm{d}\mathbf{x}_t = [ -\mathbf{f}(\mathbf{x}_t, \mathbf{y}) + g(t)^2\nabla_{\mathbf{x}_t}\log p_{t}(\mathbf{x}_t|\mathbf{y})]\mathrm{d}t + g(t)\mathrm{d}\bar{\mathbf{w}},
\end{align}
where $\nabla_{\mathbf{x}_t}\log p_{t}(\mathbf{x}_t|\mathbf{y})$ is the gradient of the conditional data probability density distribution in step $t$, also known as the \textit{score function}.
The \textit{score function} is usually unknown since the data distribution $p_{t}(\mathbf{x}_t|\mathbf{y})$ is unknown. But we can train a neural-network-based score function $\mathbf{s}_{\theta}(\mathbf{x}_t, \mathbf{y}, t)$ parameterized by $\theta$ to estimate $\nabla_{\mathbf{x}_t}\log p_{t}(\mathbf{x}_t|\mathbf{y})$ using the \textit{score matching} method \cite{hyvarinenEstimationNonnormalizedStatistical2005,vincentConnectionScoreMatching2011,songGenerativeModelingEstimating2019}. 

To train $\mathbf{s}_{\theta}$ with the \textit{score matching} method, we first sample an arbitrary time step $t$ from uniformly from $[0, T]$, and $\mathbf{x}_t$ is sampled from a Gaussian distribution of mean $\boldsymbol{\mu}(\mathbf{x}_{0}, \mathbf{y}, t)$ and variance $\sigma(t)^2$:
\begin{align}
\label{eq:score_matching_xt}
	\mathbf{x}_{t} = \boldsymbol{\mu}(\mathbf{x}_{0}, \mathbf{y}, t) + \sigma(t)\mathbf{z},
\end{align}
where $\mathbf{z} \sim \mathcal{N}_{\mathbb{R}}(\mathbf{z};0, \mathbf{I})$ , $\mathbf{I}$ is identity matrix. 
According to the solution in \cite{sarkkaAppliedStochasticDifferential2019}, the mean $\boldsymbol{\mu}$ and the variance $\sigma(t)^{2}$ at time step $t$ are deterministically computed by clean signal $\mathbf{x}_{0}$, noisy signal $\mathbf{y}$, and \textit{stiffness} $\gamma$ in Eq.\ref{eq:drift} :
\begin{align}
\label{eq:score_matching_mu}
	\boldsymbol{\mu}(\mathbf{x}_{0}, \mathbf{y}, t) = e^{-\gamma t}\mathbf{x}_{0} + (1 -  e^{-\gamma t})\mathbf{y},
\end{align}
\begin{align}
\label{eq:score_matching_sigma}
    \sigma(t)^{2} = \frac{\sigma^{2}_{\text{min}}\left( (\sigma_{\text{max}}/\sigma_{\text{min}})^{2t} - e^{-2\gamma t}  \right)\log (\sigma_{\text{max}}/\sigma_{\text{min}})}{\gamma + \log (\sigma_{\text{max}}/\sigma_{\text{min}})},
\end{align}
where $\sigma_{\text{max}}$ and $\sigma_{\text{min}}$ are hyper-parameters to control the noise scale.
After $\mathbf{x}_{t}$ and $\mathbf{z}$ in Eq.\ref{eq:score_matching_xt} are defined, the training objective 
 of \textit{score matching} can be written as:
\begin{align}
\label{eq:score_matching_loss}
\argmin_\theta \mathbb{E}_{t, (\mathbf{x}_{0}, \mathbf{y}), \mathbf{z}, \mathbf{x}_{t} | (\mathbf{x}_{0}, \mathbf{y})}\left[ \left\Vert \mathbf{s}_{\theta}(\mathbf{x}_t, \mathbf{y}, t) + \frac{\mathbf{z}}{\sigma(t)}  \right\Vert^2 \right].
\end{align}
It is noted that optimization of expectations in Eq.\ref{eq:score_matching_loss} is derived from the \textit{score matching} theory \cite{vincentConnectionScoreMatching2011}, 
even though it is very similar to training the network to estimate Gaussian noise $\mathbf{z}$ in $\mathbf{x}_{t}$ (Eq.\ref{eq:score_matching_xt}).

After the score function $\mathbf{s}_\theta$ is trained by Eq.\ref{eq:score_matching_loss}, we can replace $\nabla_{\mathbf{x}_t}\log p_{t}(\mathbf{x}_t|\mathbf{y})$ with $\mathbf{s}_\theta$ in Eq.\ref{eq:reverse_sde} for inference, and the reverse SDE can be updated as: 
\begin{align}
\label{eq:reverse_sde_theta}
    \mathrm{d}\mathbf{x}_t = [ -\mathbf{f}(\mathbf{x}_t, y) + g(t)^2\mathbf{s}_{\theta}(\mathbf{x}_t, \mathbf{y}, t)]\mathrm{d}t + g(t)\mathrm{d}\bar{\mathbf{w}}.
\end{align}

\noindent\textbf{Numerical solver:} a numerical solver can solve the continuous reverse SDE by approximating it into discrete time steps. 
The continuous interval $[0, T]$ is divided into $N$ steps with width $\Delta t = T / N$.
We apply the Predictor-Corrector samplers proposed by \cite{songScoreBasedGenerativeModeling2021a} as the numerical solver. It has two stages in each discrete time step: the predictor first updates the previous step $\mathbf{x}_{t+\Delta t}$ to the current step $\mathbf{x}_{t}$ through the reverse SDE; and the corrector refine the current $\mathbf{x}_{t}$ to make the numerical reverse process more stable.

\subsection{Score guided by discriminative models}\label{sec:discriminative_guidance}

Assume we have a pre-trained discriminative speech enhancement model $\mathcal{D}_{\phi}$ with parameters $\phi$, then we can get an enhanced signal $\mathbf{x}_{D} =  \mathcal{D}_{\phi}(\mathbf{y})$.
It can be used in place of $\mathbf{s}_{\theta}$ in Eq.\ref{eq:reverse_sde_theta} to obtain an discriminative score function at time step $t$ using: 
\begin{align}\label{eq:approx_score}
	\mathbf{s}_{D}(\mathbf{x}_{t}, \mathbf{y}, t) = \frac{\boldsymbol{\mu}(\mathbf{x}_{D}, \mathbf{y}, t) - \mathbf{x}_t}{\sigma(t)^2}.
\end{align} 
In fact, it is easy to see that, if $\mathbf{x}_{D} \to \mathbf{x}_{0}$, then Eq.~\ref{eq:approx_score} tends towards the ground-truth target scores of the \textit{score matching} method as defined in Eq. \ref{eq:score_matching_xt}-\ref{eq:score_matching_loss} above. 

Assuming that $\mathbf{x}_{D}$ is \textit{reliable enough}, we can thus replace $\mathbf{s}_{\theta}$ with $\mathbf{s}_{D}$ and warm-start the reverse diffusion sampling from a certain time-step $t=t_{\phi}$ with Eq.~\ref{eq:approx_score}; i.e. the reverse SDE in Eq. \ref{eq:reverse_sde_theta} now becomes:
\begin{align}
\label{eq:reverse_s_phi_and_theta_1}
	 \mathrm{d}\mathbf{x}_t &= [ -\mathbf{f}(\mathbf{x}_t, y) + g(t)^2\tilde{\mathbf{s}}(\cdot)]\mathrm{d}t + g(t)\mathrm{d}\bar{\mathbf{w}}, \\
  \label{eq:reverse_s_phi_and_theta_2}
	 \tilde{\mathbf{s}}(\cdot) &= \left \{ \begin{array}{lll}
	 	\mathbf{s}_{D}(\mathbf{x}_{t}, \mathbf{y}, t) & \text{if} & t > t_{\phi}, \\
	 	\mathbf{s}_{\theta}(\mathbf{x}_t, \mathbf{y}, t) & \text{if} & t \leq t_{\phi}.
	 \end{array}
	 \right.
\end{align}
As such, during the reverse-diffusion process, the number of discrete steps that use $\mathbf{s}_{D}$ as the score will be $N_{\phi} = |\{\Delta t \times n | \Delta t \times n > t_{\phi}\}_{n=1}^{N}|$.
The greater $N_{\phi}$, the fewer reverse diffusion steps with the score-matching-trained model $\mathbf{s}_{\theta}$ are required, leading to significantly reduced computational overhead. On the other hand, a large $N_{\phi}$ will lead the reverse diffusion process to converge to $\mathbf{x}_{D}$.
These aspects will be explored in Section ~\ref{sec:results}. 

Note that our approach only takes effect in the inference stage.
We do \textbf{not} need to modify the training of the SGM $\mathbf{s}_{\theta}$ to get the guidance from $\mathcal{D}_{\phi}$. 
It fundamentally differs from a recent work named StoRM \cite{lemercierStoRMDiffusionBasedStochastic2023}.  
In StoRM, $\mathbf{s}_{\theta}$ is trained to refine the output of a discriminative model, and this latter is actively employed in the training procedure. 
Our approach is instead totally agnostic to the choice of $\mathcal{D}_{\phi}$ and allows for more flexibility in inference.

\begin{table*}[thb]
\caption{Comparison of online/offline and generative/discriminative models on WSJ0-CHiME3 data; \textbf{-D} for discriminative models; \textbf{-G} for generative models; ``4.a+3'' in Sys. ID means ``Sys.4.a'' guided by ``Sys.3'' ; 
Metrics higher is better except for MCD and LSD.}

\label{tab:online_wsj0_chime3}
\renewcommand{\arraystretch}{0.90} 
\setlength\tabcolsep{2pt}
\begin{tabular}{@{}cccccc|ccccccc|cc@{}}
\toprule
 &  &  &  &  & &  \multicolumn{7}{c|}{\textbf{WSJ0-CHiME3}} & \multicolumn{2}{c}{\textbf{CHiME4 Real}} \\
\multirow{-2}{*}{\textbf{Sys.ID}} & \multirow{-2}{*}{\textbf{Model}} & \multirow{-2}{*}{\textbf{\begin{tabular}[c]{@{}c@{}}Ideal\\ Latency\end{tabular}}} & \multicolumn{1}{c}{\multirow{-2}{*}{\textbf{\begin{tabular}[c]{@{}c@{}}\# Para.\\ (M)\end{tabular}}}} & \multirow{-2}{*}{$\boldsymbol{N}$} & \multirow{-2}{*}{$\boldsymbol{N_{\phi}}$} &\textbf{\begin{tabular}[c]{@{}c@{}}DNSMOS \\ OVRL\end{tabular}} & \textbf{\begin{tabular}[c]{@{}c@{}}NISQA \\ OVRL\end{tabular}} & \textbf{PESQ} & \textbf{ESTOI} & \textbf{\begin{tabular}[c]{@{}c@{}}SDR \\ (dB)\end{tabular}} & \textbf{LSD} $\downarrow$ & \textbf{MCD} $\downarrow$ & \textbf{\begin{tabular}[c]{@{}c@{}}DNSMOS \\ OVRL\end{tabular}} & \textbf{\begin{tabular}[c]{@{}c@{}}NISQA \\ OVRL\end{tabular}}  \\ \midrule
- & Noisy speech  & -  & - &- & - &  2.72 & 2.70 & 1.69 & 0.78 & 9.69  & 3.00 & 5.43 & 1.39 & 1.10 \\

- & SGMSE+ \cite{richterSpeechEnhancementDereverberation2023a} & offline & 66 & - & - & - & - & 2.96 & 0.92 & 18.30 & - & - & - & -\\
0 & SGMSE+* & offline & 66 & 30& 0 &  3.31 & 4.22 & 2.84 & 0.92 & 17.69 & 1.89 & 1.40 & 2.88 & 3.89 \\
\rowcolor[HTML]{EFEFEF} 
1 & SkiM-D & offline & 35 & - & - &  \textbf{3.38} & \textbf{4.29} & \textbf{3.07} & \textbf{0.94} & \textbf{20.80} & \textbf{1.87} & \textbf{1.10} & 2.94 & 2.98 \\
2 & SkiM-G & offline & 37 & 30 & 0 &  3.15 & 3.91 & 2.46 & 0.90 & 16.76 & 2.46 &  2.30 & \textbf{3.15} & \textbf{3.92} \\ 
\rowcolor[HTML]{EFEFEF} 
\midrule
\rowcolor[HTML]{EFEFEF} 
3 & SkiM-D & 50ms & 8 & - & - & {3.18} & 3.50 & \textbf{2.78} & \textbf{0.93} & 20.02 & \textbf{1.90} & \textbf{1.24}  & 2.26 & 1.86 \\
4.a & SkiM-G & 50ms & 46 & 30 & 0 & 3.03 & 3.82 & 2.46 & 0.90 & 16.88 & 2.49 & 2.53 & 2.24 & 2.18\\ 
4.b & SkiM-G & 50ms & 46 & 15 & 0 & 2.18	& 1.81 & 1.22 & 0.66 & 5.38 & 5.09 & 7.52 & 1.96	& 1.47 \\


\midrule
4.a+3 & SkiM-G & 50ms & 46 + 8 & 30 & 12 & 3.06 & \textbf{3.86} & 2.51 & 0.91 & 17.47 & 2.46 & 2.34 & {2.30} & \textbf{2.26} \\
4.b+3 & SkiM-G & 50ms & 46 + 8 & 15 & 13 & \textbf{3.19} & 3.81 & 2.61 & 0.92 & 19.38 & 2.10 & 1.47 & \textbf{2.36} & 1.98 \\
\bottomrule
\end{tabular}
\end{table*}

\subsection{Streamable chunk-Level diffusion-based SE} 
\label{sec:streamable_sgmse}

To make SE with DGMs streamable, we further propose a novel chunk-based diffusion architecture. It can be applied to causal DNNs with states that encode the history states, e.g., recurrent neural networks (RNNs).

In the training stage, we follow the Eq. \ref{eq:score_matching_loss} with the training pair $\mathbf{y} \in \mathbb{R}^{1\times L}$ and $\mathbf{x}_{0} \in \mathbb{R}^{1\times L}$ of length $L$.
In the inference stage, we implement a chunk-level streaming inference method.
Denote $y^{c} \in \mathbb{R}^{1\times K} $ as a small chunk of the noisy speech signal, where $K$ is the chunk size ($50$\,ms in this paper), as well as the algorithmic latency of the online model.
It is noted that $c$ indicates $c$-th chunk in the order of time, while $t$ indicates the step in the diffusion process.
Then, the reverse SDE in the sampling stage is:
\begin{align}
\label{eq:streaming}
    \mathrm{d}\mathbf{x}_t^{c} = [ -\mathbf{f}(\mathbf{x}_t^{c}, \mathbf{y}^{c}) + g(t)^2\mathbf{s}_{\theta}(\mathbf{x}_t^{c}, \mathbf{y}^{c}, t, \mathcal{H}_{t}^{c})]\mathrm{d}t + g(t)\mathrm{d}\bar{\mathbf{w}}^{c}, 
\end{align}
where $\mathcal{H}_{t}^{c}$ is a set of history states which encode the information from the $1$st streaming chunk to the $(c-1)$th chunk at diffusion step $t$. 
During the streaming inference, $\{\mathcal{H}_{t}^{c}\}_{t=0}^{T}$ can be maintained to store the memory of all the processed history and help to process the current chunk. 
In other words, the reverse diffusion process is run for each chunk of the input signal and is conditioned by the history of the previously processed chunks.

\section{Experimental Setups}

\subsection{Dataset}

\noindent \textbf{Training.} We use the WSJ0-CHiME3 dataset created by \cite{richterSpeechEnhancementDereverberation2023a}. It is a simulated dataset for speech enhancement. The clean speech signals come from the WSJ0 dataset \cite{garofolojohns.CSRIWSJ0Complete2007}, and the noisy signals are from the CHiME3 challenge dataset \cite{barkerThirdCHiMESpeech2015}. We use the recipe\footnote{https://github.com/sp-uhh/sgmse} provided by \cite{richterSpeechEnhancementDereverberation2023a} for data simulation.

\noindent \textbf{Testing.} We adopt two different testing datasets for evaluation.
The first testing set is WSJ0-CHiME3, which is domain-matched with the training set.
The second testing set is the real data from the CHiME4 dataset \cite{barkerThirdCHiMESpeech2015,vincentAnalysisEnvironmentMicrophone2017}. Instead, the real CHiME4 is domain-mismatched and, thus, used to assess generalizability. 

\subsection{Evaluation metrics}
To evaluate the performance of the proposed approach from multiple perspectives, we use several metrics: perceptual evaluation of speech quality (PESQ) \cite{rixPerceptualEvaluationSpeech2001}, extended short-time objective intelligibility (ESTOI) \cite{jensenAlgorithmPredictingIntelligibility2016}, signal-to-distortion ratio (SDR) \cite{vincentPerformanceMeasurementBlind2006}, log-spectral distance (LSD) \cite{grayDistanceMeasuresSpeech1976}, mel cepstral distortion (MSD) \cite{kubichekMelcepstralDistanceMeasure1993}, NISQA \cite{mittagNISQADeepCNNSelfAttention2021}, and overall DNSMOS(P.835) score \cite{reddyDnsmos835NonIntrusive2022}  as evaluation metrics. 

\subsection{Model configurations}
\noindent \textbf{Neural Networks.}
We first reproduce the SGMSE+ model \cite{richterSpeechEnhancementDereverberation2023a} as one of the baselines with the ESPNet-SE \cite{liESPnetSEEndToEndSpeech2021,luESPnetSESpeechEnhancement2022} toolkit. We follow the configurations in \cite{richterSpeechEnhancementDereverberation2023a} for the SGMSE+, except that we did not use the exponential moving average in training. Instead, for the model used in this work, we store the 5 lowest-loss model parameters and average them to obtain the final model for inference.

Following the chunk-level streamable diffusion framework described in Section~\ref{sec:streamable_sgmse}, we use SkiM \cite{liSkimSkippingMemory2022} as the score model for online diffusion models.
SkiM consists uses two different long short-term memory (LSTM) blocks: one for local short-segment processing (segment-LSTM) and another long-span memory maintenance (memory-LSTM).
Originally, in SkiM~\cite{liSkimSkippingMemory2022,liPredictiveSkimContrastive2023}, LSTMs in both blocks are uni-directional to minimize the ideal latency for low-latency, frame-level streaming.
However, to make the diffusion reverse process more stable, we make the segment-LSTM bi-directional and keep the memory-LSTM unidirectional, which leads to a chunk-level streaming model.
For all the diffusion SkiM models, we use the model with $8$ basic SkiM blocks \cite{liSkimSkippingMemory2022}, and the hidden dimension width is 256 for uni-direction memory-LSTMs and 512 in-total for bi-direction segment-LSTMs. 
We also build discriminative SkiM enhancement models for comparison and to also play the part of $\mathcal{D}_{\phi}$ for our discriminative guidance experiments.
The signal-to-noise ratio (SNR) loss is used to train discriminative models.
We train all the models for a max of 100 epochs on the WSJ0-CHiME3 training set.

\noindent \textbf{Diffusion.}
We set $\gamma$ in Eq.~\ref{eq:drift} to $\gamma = 1.5$; $T = 1.0$; We use $\sigma_{\text{max}} = 0.5 $, and $ \sigma_{\text{min}} = 0.05$ for the complex domain SGMSE+, while  $\sigma_{\text{max}} = 10^{-1} $, $ \sigma_{\text{min}} = 10^{-4}$ for the time domain SkiM online models.
In the Predictor-Corrector sampling,  we adopt the reverse diffusion SDE method \cite{songScoreBasedGenerativeModeling2021a} as the Predictor and the annealed Langevin Corrector \cite{songScoreBasedGenerativeModeling2021a}.

\begin{figure*}[t]
\centering
\includegraphics[width=0.95\linewidth]{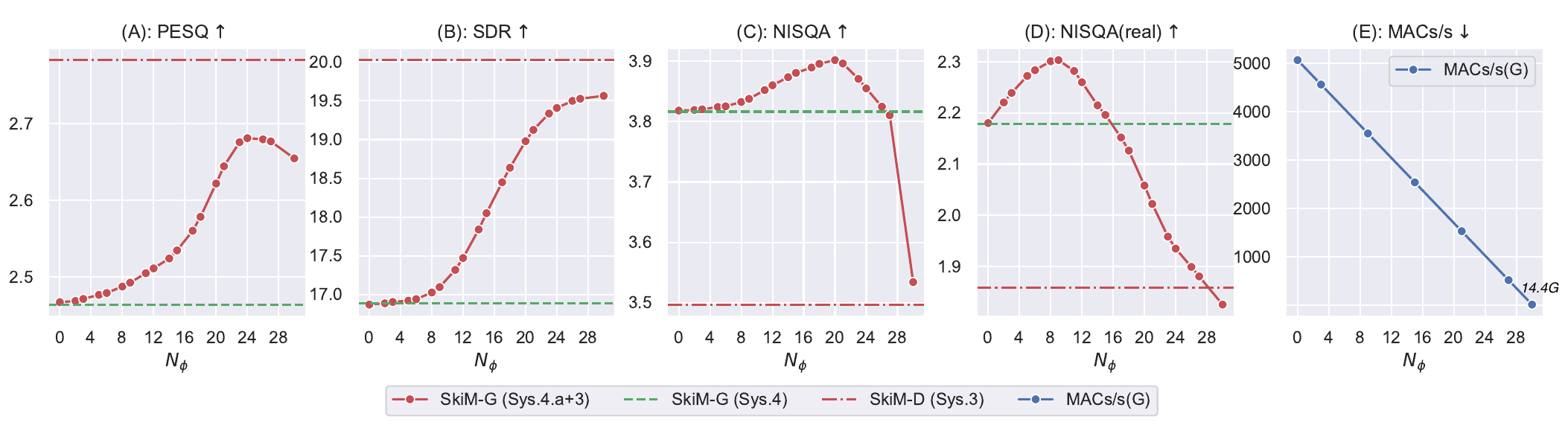}
\caption{(A)-(D): Effect of proposed discriminative guidance on WSJ0-CHiME3 and CHiME4(real) test set. The total number of reverse steps is $N=30$, $\mathbf{s}_{\theta}$ is replaced by $\mathbf{s}_{D}$ in the first $N_{\phi}$ steps of $N$; (E): GMACs/s as $N_{\phi}$ increases; the 14.4 GMACs/s for $N_{\phi}=30$ corresponds to using only SkiM-D (Sys.3).}
\label{fig:fig_1}
\end{figure*}

\section{Results and Discussions}
\label{sec:results}

\subsection{Results of online SGMSE and algorithm latency}

Table~\ref{tab:online_wsj0_chime3} compares the performance between online and offline SGMSE models. 
Top panel reports offline baseline systems: SGMSE+ model\cite{richterSpeechEnhancementDereverberation2023a} (Sys.0), a generative SkiM (SkiM-G, Sys.2) and a discriminative SkiM (SkiM-D, Sys.1). 
For these offline models, we can observe that SkiM-D, compared to SkiM-G, performs better overall on the fully matched WSJ0-CHiME3 test set. However, SkiM-G shows better generalization ability on the CHiME4 real test-set, thus confirming the findings in~\cite{luConditionalDiffusionProbabilistic2022}.

Mid panel reports several online models with 50\,ms ideal latency (corresponding to chunk size $K=800$ in Section~\ref{sec:streamable_sgmse}), both generative (Sys.\{4.a,4.b\}) and discriminative (Sys.3).
Remarkably, by comparing the performance of the online systems Sys.\{3,4a\} with their respective offline counterpart Sys.\{1,2\}, we observe no significant performance loss on the WSJ0-CHiME3 test set. However, DNSMOS and NISQA scores of the online systems are significantly worse in the CHiME4 scenario, suggesting that generalization is more difficult in the online scenario. 
Comparing online SkiM-G (Sys.4a) with its respective discriminative counterpart (Sys.3), the trend observed for offline systems remains: SkiM-G has better NISQA scores on both test sets but is inferior regarding the other metrics, which are intrusive and thus may penalize more generative approaches. 
We also report results with SkiM-G by halving the reverse diffusion steps (Sys.4b) from $30$ to $15$. 
This leads to a drastic reduction in performance even if the computational requirements would be halved. 







\subsection{Results with discriminative guidance}

In the bottom panel of Table~\ref{tab:online_wsj0_chime3}, we report the effect of our proposed discriminative guidance on the $50$-ms online generative models.
In detail, we report results for both SkiM-G with $N=30$ (Sys.4a+3) and with $N=15$ (Sys.4b+3) numbers of total reverse diffusion steps. For these systems we used the online SkiM-D (Sys.3) as the discriminative model $\mathcal{D}_{\phi}$ to obtain $\mathbf{s}_{D}$ for the first  $N_{\phi}$ steps.
For both of these, we report the choice of $N_{\phi}$ that maximizes the overall performance. 
We can observe that, in both instances, discriminative guidance helps considerably, especially regarding generalization to the real-world CHiME4 test set. 
The two guided systems appear to have the best of both worlds: both desirable generalization and higher non-intrusive perceptual metrics but also high intrusive metrics thanks to the discriminative guidance. 

Remarkably, in both cases, this also comes with a significant reduction in computational requirements as the diffusion steps with $\mathbf{s}_{D}$ are drastically reduced. 
The results with $N=15$ steps (Sys.4b) are especially promising as we show that even just two reverse diffusion steps with $\mathbf{s}_{D}$ can be enough (again,  $\mathbf{s}_{D}$ needs only one forward from $\mathcal{D}_{\phi}$ for all $N_\phi$).

In Figure~\ref{fig:fig_1}, we report the trend on $3$ representative metrics (PESQ, SDR, NISQA) with the discriminative guidance system (Sys.4a+3) as the number of $N_{\phi}$ increases and approaches the number of total steps $N=30$. 
The constant lines are, respectively, the pure discriminative model (Sys.3, red) and the pure generative model (Sys.4a, green). Note that this latter, (Sys.4a, green), also corresponds to the $N_{\phi}=0$ case.
As $N_{\phi}$ increases, we can observe that the system with the guidance deviates from the pure generative model performance. 
On the intrusive metrics (PESQ, SDR), the guided model tends to the discriminative model performance (which is higher on these). Regarding NISQA, the guided model can even increase performance with respect to the pure generative model ($N_{\phi}=0$), at least up to a certain number of $N_{\phi}$ steps. For higher $N_{\phi}$ it converges to the NISQA obtained with the discriminative model alone. 
This suggests that $N_{\phi}$ can be used to trade-off between the discriminative and generative capabilities of the overall system.


Finally, we also report the number of the multiplier–accumulator (MACs) operations for per second audio with the \textit{thop} \footnote{\url{https://github.com/Lyken17/pytorch-OpCounter}} toolkit.
Fig.\ref{fig:fig_1}.E shows MACs/s in the inference stage with respect to $N_{\phi}$, with the total reverse step $N = 30$. 
The computational cost from the discriminative model (Sys.3) is included in the figure. 
It is easy to see that the computational cost descends linearly by increasing $N_{\phi}$, since we do not need a DNN forward for each step to re-estimate  $\mathbf{x}_{D}$ to obtain $\mathbf{s}_{D}$ with Eq.\ref{eq:approx_score}.

\section{Conclusion and Future Works}

In this work, we propose a new streamable diffusion method for speech enhancement.
The reverse diffusion process is performed at the chunk level, and the history of the previously processed chunks is maintained and leveraged to help denoise the current chunk. 
To decrease the computational footprint, we devised a new framework in which the score function can be estimated with the guidance of discriminative models.
We can significantly reduce the number of DNN forward operation in the reverse process by using the guided score without sacrificing performance.
Future work will focus on 1) how latency can be further reduced, 2) whether this discriminative guidance framework can be leveraged for other applications, and 3) exploring more strategies to balance generative scores and discriminative guidance.


\ifinterspeechfinal
\section{Acknowledgment}
The experiments were done using the PSC Bridges2 system via ACCESS allocation CIS210014, supported by National Science Foundation grants \#2138259, \#2138286, \#2138307, \#2137603, and \#2138296.
\fi

\bibliographystyle{IEEEtran}
\bibliography{mybib}

\end{document}